\documentclass[11pt]{article}
\usepackage{hyperref}
\usepackage{amssymb}
\usepackage{graphicx}
\usepackage{amsmath}
\usepackage{authblk}
\usepackage{cite}
\usepackage{physics}
\usepackage{revsymb}
\usepackage{float}
\usepackage{slashed}
\usepackage{tabularx}
\usepackage{breqn}
\usepackage{array}
\usepackage{verbatim}
\usepackage{dsfont}
\usepackage[section]{placeins}
\usepackage[a4paper, total={6.8in, 10in}]{geometry}
\numberwithin{equation}{section}
\newcommand{\be}{\begin{equation}}
\newcommand{\ee}{\end{equation}}
\newcommand{\bea}{\begin{eqnarray}}
\newcommand{\eea}{\end{eqnarray}}
\newcommand{\ba}{\begin{aligned}}
\newcommand{\ea}{\end{aligned}}

\begin{document}
\title{Yang-Mills Field in the $\kappa$-space-time}
\author{ {\bf {\normalsize Bhagya. R}\thanks{22phph03@uohyd.ac.in, r.bhagya1999@gmail.com},\,
{\bf {\normalsize E. Harikumar}\thanks{eharikumar@uohyd.ac.in}}}\\
{\normalsize School of Physics, University of Hyderabad}\\{\normalsize Central University P.O, Hyderabad-500046, Telangana, India}\\ 
}
\date{}
\maketitle
\begin{abstract}
In this paper, we construct $SU(N)$ Yang-Mills theory in the $\kappa$-space-time, valid up to first order in the deformation parameter $a$, using the generalisation of Feynman's approach. Using the $\kappa$-deformed Wong's equation derived, in the Jacobi identity involving velocities and coordinates of $\kappa$-deformed space-time, the $\kappa$-deformed homogeneous Yang-Mills equations are derived. We show the compatibility between the $\kappa$-deformed field strength derived using the Jacobi identity and the commutators of the gauge covariant derivative, up to first order in $a$. The $\kappa$-deformed field strength is covariant under $SU(N)$ gauge transformations. We then construct the Lagrangian for Yang-Mills theory in $\kappa$-deformed space-time and show that it is invariant under $SU(N)$ transformation and not under U(N) transformation. We also derive the expression for the force experienced by an isospin-carrying particle in the presence of Yang-Mills field in the $\kappa$-space-time.
\end{abstract}

\section{Introduction}
 The microscopic theory of gravity requires the amalgamation of general relativity and quantum mechanics, and such a theory is needed to understand and explain the nature of gravitational force at very short distances, well below nuclear size. Though many different approaches have been developed to discuss quantum gravity, with varying degrees of success, a completely satisfactory quantum theory of gravity still remains elusive. But one of the common features all these approaches share is the existence of a fundamental length scale; it is below this length scale, the quantum effects of gravity become important. As non commutative(NC) space-time(s) inherently introduce a length scale, non commutative geometry promises to be a viable framework for consistent quantisation of gravity \cite{connes}. Thus, it is also interesting to construct and study field theory models on non commutative (NC) space-times and analyse their quantisation. In particular, construction of the standard model in NC space-time is of immense importance, and this was carried out in \cite{connes,landi}. These studies naturally led to the construction and investigation of gauge theories on NC space-time. NC standard model along with a consistent quantum gravity also opens up the possibility of grand unified models \cite{connes, landi}.

As the existence of a fundamental length scale is not compatible with the special theory of relativity, the principle of relativity needed to be modified. This leads to doubly special relativity (DSR), which admits a fundamental length scale and speed of light as universal constants \cite{DSR}. The underlying space-time of DSR is known to be $\kappa$-space-time \cite{Ruegg, Ruegg1, majid}. This $\kappa$-space-time also appears in the low-energy limit of quantum gravity models \cite{freidel}. The symmetries of $\kappa$-space-time are realised as Hopf algebra \cite{Ruegg, Ruegg1}. Various issues in constructing quantum field theories in $\kappa$-space-time are addressed and resolved in \cite{Ruegg, Ruegg1, majid,freidel,lukierski2,kosinski,kosinski1,kosinski2,amelino, wess,jonke3,jonke, MD jonke, jonke2, MD jonke1,jonke1,tim,tim2}.

The present status of research on building gauge theories in various NC space-times, as well as attempts to include gravity into the NC framework, is discussed in \cite{hersent1}.
 Deformed gauge theories are constructed for arbitrary compact Lie groups, and the Lagrangian for such gauge theories possessing a deformed Lorentz covariance is constructed in \cite{wess}. The symmetries of $\kappa$-space-time are realised using undeformed $\kappa$-Poincare algebra \cite{SM}. Here, different possible realisations of space-time coordinates of $\kappa$-space-time in terms of the commutative variables are introduced.

The studies of quantum field theories in $\kappa$-space time are carried out in \cite{twisted, TRG, sivakumar, EH, ED, geodesic, wallet, vitale2, wallet2, marmo, zampini, dvv, zet, tureanu, schraml,  lizzi, vitale, kurkov, brunner, kupriyanov, kontsevich, LP, PD}. In \cite{twisted, TRG,sivakumar}, $\kappa$-deformed space-time coordinates were written in terms of commutative coordinates and their derivatives \cite{SM}. This provided a systematic way of writing down field theory models in $\kappa$-space-time in terms of commutative variables. This approach has the advantage of applying well-established field theoretical tools of commutative space-time in analysing the effects of $\kappa$-deformation and is known to be equivalent to the star product approach \cite{SM}, where one replaces point-wise product by a modified multiplication rule known as star product.


In \cite{dyson}, two of the Maxwell's equations were derived by assuming Newton's law and commutation relations between position coordinates and corresponding velocities. The remaining equations were argued to be defining the sources of electric and magnetic fields, respectively, \cite{dyson}. This was followed by the derivation of the Yang-Mills equation by considering a non-relativistic particle carrying isospin \cite{lee}. In \cite {tanimura}, a covariant generalisation of the derivation of the Maxwell's equations and the Yang-Mills equation was developed. The construction of the Yang-Mills equation in this approach crucially depends on Wong's equation describing the motion of a particle carrying isospin charge, in the presence of an external non-abelian gauge field \cite{wongs}. Maxwell's equation in $\kappa$-space-time was derived by generalising  Feynman's approach in \cite{EH, ED}.  Generalising this minimal coupling prescription \cite{ED} to curved, $\kappa$-deformed space-time, the geodesic equation in $\kappa$-deformed space-time was derived and analysed in \cite{geodesic}.

Construction and analysis of non-abelian gauge theories in $\kappa$-space-time is an active area of research \cite{kosinski, kosinski1, kosinski2, amelino, wess,jonke3, jonke, MD jonke, jonke2, MD jonke1, jonke1, tim, tim2, hersent1, SM, twisted, TRG, sivakumar, EH, ED, geodesic, wallet, vitale2, wallet2, marmo, zampini, dvv, zet, tureanu, schraml, lizzi, vitale,kurkov, brunner,kupriyanov, kontsevich, LP, PD}. In \cite{jonke3, jonke}, a general formalism for constructing field theories on quantum space-time is developed, and it is used to obtain field theory models in $\kappa$-space-time. This was followed by the construction of U(1) gauge theory in $\kappa$-space-time\cite{MD jonke, jonke2}.

In \cite{wallet,vitale2,wallet2,marmo,zampini,dvv,zet,tureanu,schraml,lizzi,vitale}, NC differential calculus was developed and used to construct NC field theories. Field theory models constructed on NC space-times with NC parameter depending on the space-time coordinates using the Kontsevich star product \cite{kontsevich} were shown to have inconsistent commutative limit. This difficulty was resolved in \cite{vitale}, wherein the bootstrap method was used.\cite{brunner, kupriyanov}. All these results are valid up to the first order in the NC parameter. In \cite{kurkov}, modified action for a $U(1)$ model on $\kappa$-space-time was constructed. It was shown in \cite{LP}, that the action obtained in \cite{kurkov} does not have the correct commutative limit. Gauge covariant equations of motion were constructed ({which are {\it not Euler-Lagrange equation}), and its solutions were obtained
in \cite{kurkov}. It was shown that this $U(1)$ theory is invariant under a modified gauge transformation.
In all these studies, the gauge theory models constructed are mostly valid up to first order in NC parameter, and {\it (1) are invariant under either $U(1)$ or $U(N)$ gauge transformations, (2) some of these models do not have correct commutative limit}. In this work, using the covariant form of the Feynman's approach, we construct $\kappa$-deformed gauge theory valid up to first order in the NC parameter,
\begin{itemize}
 \item  which is invariant under $SU(N)$ transformations, and
 \item  reduces to usual SU(N) invariant theory in the commutative limit
\end{itemize}
and we also obtain the corresponding $\kappa$-deformed Yang-Mills Lagrangian. The $\kappa$-deformed Yang-Mills equation is derived by exploiting Jacobi identities involving various combinations of $\kappa$-deformed coordinates and corresponding velocities. In this approach, taking into account of the symmetry property of the commutators between velocities, we introduce two second rank tensors, one of which is anti-symmetric. This anti-symmetric tensor is later shown to be the Yang-Mills field strength. The Jacobi identities then show that this Yang-Mills field strength satisfies homogeneous Yang-Mills equations. In evaluating the Jacobi identities, one needs the commutation relations between $\kappa$-deformed velocities and Lie algebra-valued functions defined on $\kappa$-space-time. This naturally requires the commutation relation between different components of the $\kappa$-deformed velocities with generators of the Lie algebra and their time derivatives. The Wong's equation, describing the motion of isospin carrying particle in the background of non-abelian gauge field, is used in the derivation of the Yang-Mills equation at this stage. The Wong's equation comes with a condition on the time development of the generators of the Lie algebra associated with the non-abelian gauge field. This condition plays an important role in the derivation of the Yang-Mills field equation using Feynman's approach. We see that the gauge covariant derivatives and hence non-abelian ``electric'' and ``magnetic'' fields in the $\kappa$-space time are modified in the same general form but different in the numerical factors
and their commutators reproduce the components of the $\kappa$-deformed field strength. Using the $\kappa$-deformed field strength, we then write down the Lagrangian describing $\kappa$-Yang-Mills field theory. The remaining Yang-Mills equations follow as Euler-Lagrange equations from this $\kappa$-deformed Lagrangian. The force equation for a particle carrying isospin, interacting with Yang-Mills field, is then derived. All our results are valid up to first order in the deformation parameter $a$.
This $SU(N)$ Yang-Mills theory in $\kappa$-space-time constructed here is a first step in developing the standard model in $\kappa$-space-time with the same gauge group as in the commutative space time.
 
  After quickly recalling the derivation of Maxwell's equations and Yang-Mills equation using Feynman's approach, and summarising the essentials of undeformed $\kappa$-Poincare algebra, in the next section, in section 3, we present the derivation of $\kappa$-deformed Wong's equation. We then derive, in section 4, the Yang-Mills equation in the $\kappa$-deformed space-time. Commutator between the $\kappa$-deformed velocity and Lie algebra valued functions of $\kappa$-deformed space-time is an integral part of this derivation and is calculated using $\kappa$-deformed Wong's equation. Using the explicit form of the Yang-Mills field strength in the $\kappa$-deformed space-time obtained, we also construct the Lagrangian describing $\kappa$-deformed Yang-Mills theory in section 4. We show that this model is invariant under $SU(N)$ transformations. We then derive the force experienced by an isospin carrying particle moving in the presence of Yang-Mills fields in section 5. Our concluding remarks are given in section 6.

\section{Preliminaries: Feynman's approach, $\kappa$-space-time}
Here, we summarise Feynman's approach of deriving gauge field equations first. This is followed by a quick summary of the realisation approach used in studying $\kappa$-models.

\subsection{Gauge field equations from Newtons's Law: Feynman's approach}

In Feynman's approach \cite{dyson}, Maxwell's equations are derived by assuming Newton's law and the commutation relations between position coordinates and velocities, i.e.,
\bea \label{NF}
m\ddot{x}_j = F_j (x, \dot{x},t),
\eea
\be \label{commutator}
[x_i,x_j] = 0,~~~m[x_j,\dot{x}_k] = i\hbar \delta_{jk}.
\ee
Using the time derivative of the commutator given above and the Jacobi identity between $x_i$ and velocities, one finds $[x_i, F_j]$ to be anti-symmetric in indices. Thus, an anti-symmetric tensor is introduced as
\be \label{MF}
[x_j,F_k] = -\frac{i \hbar e}{m} \epsilon_{jkl} B_l.
\ee
Considering eq.(\ref{MF}) as the definition of B and using the Jacobi identity, it is shown that B is a function of x and t only. Use of eq.(\ref{commutator}) in eq.(\ref{MF}) allows one to find $F_k$ by integrating right hand side of eq.(\ref{MF}) with respect to $\dot{x}_k$ setting
\be \label{force}
F_j = e E_j+ e\epsilon_{jkl}\dot{x}_k B_l,
\ee
which is the Lorentz force equation for a particle of charge $e$.
Taking eq.(\ref{force}) as the definition of the field E, we see $[x_j,E_k] =0$.
Rewriting eq.(\ref{MF}) as 
\be \label{Hdef}
B_l = - \frac{im^2}{2e \hbar}\epsilon_{ljk}[\dot{x}_j, \dot{x}_k]
\ee 
and applying Jacobi identity involving velocities, we get
\be
[\dot{x}_l,B_l] =0,
\ee
which leads to
\be
\partial_i B_i =0.
\ee
By taking the total time derivative of the definition of $B$ given in eq.(\ref{Hdef}), the second Maxwell's equation is obtained as
\be
\frac{\partial B_l}{\partial t} = \epsilon_{ljk} \frac{\partial E_j}{\partial x_k}. 
\ee
The other two Maxwell's equations define the charge and current density.

In \cite{tanimura}, Feynman's approach is generalised to the relativistic case and extended to non-abelian gauge theory, generalising the results of \cite{lee,dyson}.
In \cite{tanimura}, a particle carrying isospin $I^a$ satisfying,
\be \label{commutator1}
m[x_\mu, \dot{x}_\nu] = -i\hbar \eta_{\mu \nu},~~\eta_{\mu \nu} ={\rm diag} (1 , -1, -1 ,-1),
\ee
\be
[I^a,I^b]=i \hbar f^{abc}I^c,~~~ [x^\mu, I^a] =0,
\ee
and the Wong's equation
\be
\dot{I}^a - ef^{abc}A^b_\mu \dot{x}^\mu I^c =0,
\ee
is considered. In the above equation, an overdot stands for derivative with respect to $\tau$, which is the time parameter introduced in the covariant generalisation of Feynman's approach \cite{tanimura}.
Here one follows the same prescription as in \cite{dyson}, starting by differentiating eq.(\ref{commutator1}) with respect to $\tau$ and define,
\be \label{tensor1}
F^{\mu \nu} =\frac{m}{i \hbar e}[x^\mu, F^\nu] = -\frac{m^2}{i \hbar e} [\dot{x}^\mu, \dot{x}^\nu],
\ee
where one used $F^\nu = m \ddot{x}^\nu$.
By using Jacobi identity, it is shown that $F_{\mu \nu}$ is a function of $x^\mu$ alone, which allows the integration of the above expression, resulting in 
\be
F^\mu (x, \dot{x},t) = G^{a\mu}(x)I^a+ <e F^{a\mu}_{\nu} I^a\dot{x}^\nu>.
\ee
Using Wong's equation, the commutator of four-velocities with Lie algebra-valued functions is calculated. This also provides us with the definition of gauge covariant derivatives. Then by using the Jacobi identity involving velocities and the definition of $F^{\mu \nu}$ given in eq.(\ref{tensor1}), one obtains
\be \label{commutativeequation}
D_\mu F_{\nu \rho} + D_\nu F_{\rho \mu} + D_{\rho} F_{\mu \nu} =0.
\ee
where $(D_{\mu}\phi)^a = \partial_{\mu} \phi^a
-e f^{abc} A^b_{\mu} \phi^c$ is the gauge covariant derivative. Also by using the commutation relation $[\dot{x}_\mu, I^a]$, obtained using Wong's equation and Jacobi identities one finds the explicit expression for $F^a_{\mu \nu}$ in terms of gauge fields as  $F^a_{\mu \nu} = \partial_\mu A^a_\nu - \partial_\nu A^a_\mu -ef^{abc}A^b_\mu A^c_\nu$. In this approach, the homogeneous Yang-Mills equations and force equation are obtained.
In our study, we generalise this approach to the $SU(N)$ gauge theory in the $\kappa$-deformed space-time.

\subsection{Undeformed $\kappa$ Poincare algebra: essentials}

Instead of using differential calculus on non commutative space-time, one can rewrite the non commutative coordinates in terms of the commutative coordinates and usual derivatives, in such a way that the commutation relation between non commutative coordinates is satisfied \cite{SM}. This allows one to set up models on non commutative space-time in terms of commutative coordinates, their derivatives, and functions of these commutative variables. In this approach, the effect of non commutativity enters through the non commutative parameter-dependent terms that appear in the expressions of non commutative coordinates. We employ this approach as it allows the use of tools of commutative space-time to construct and analyse the model, and also makes the comparison with the commutative model easy. Thus, we start with

\be
\hat{x}_0 = x_0 \psi(ia\partial_0) + ia x_j \partial_j\gamma(ia\partial_0); ~~~~~
\hat{x}_i = x_i \varphi(ia\partial_0)
\ee
Imposing the condition that this should satisfy
\be \label{kappa1}
[\hat{x}_{\mu}, \hat{x}_{\nu}] = i (a_{\mu} \hat{x}_{\nu} - a_{\nu} \hat{x}_{\mu}),
\ee
with the NC deformation parameter $a_0=a=\frac{1}{\kappa}$, $a_i =0$ and demanding these $\hat{x}_0$ and $\hat{x}_i$ to have correct commutative limit leads to the conditions
\bea
\frac{1}{\varphi(ia\partial_0)} \frac{d \varphi}{d(ia\partial_0)} \psi(ia\partial_0) = \gamma(ia\partial_0) -1, \\
\psi(0)=1;~~~~ \varphi(0) =1,
\eea
respectively.
Here for $\psi$ and $\varphi$ different choices are allowed and we choose $\psi(ia\partial_0) =1$ and $\varphi(ia\partial_0) = e^{-ia\partial_0}$ in our study \cite{SM}. Thus we have
\be \label{Rkappa}
\hat{x}_0 =x_0;~~~~ \hat{x}_i = x_i e^{-ia\partial_0}
\ee
In this realisation approach, the associated symmetry algebra of the $\kappa$-Minkowski space-time is the undeformed $\kappa$-Poincare algebra \cite{SM}, whose generators are modified, whereas the defining relations are exactly the same as the usual Poincare algebra.
The corresponding generators are given as \cite{SM}
\bea
M_{ij} &=& x_i \partial_j -x_j \partial_i \\ \nonumber
M_{i0} &=& x_i \partial_0 \varphi (\frac{e^{2ia\partial_0}-1}{2ia\partial_0}) -x_0 \partial_i \frac{1}{\varphi} + ia x_i \partial^2_k \frac{1}{2 \varphi} -iax_k\partial_k \partial_i\frac{\gamma}{\varphi}
\eea
Since derivative $\partial_\mu$ is not invariant under undeformed $\kappa$-Poincare algebra, they are replaced by the Dirac derivative $D_\mu$. Under this algebra, the Dirac derivative transforms as a 4-vector,
\bea \label{generator}
[M_{\alpha \beta}, D_\mu] = \eta_{\beta \mu} D_\alpha  - \eta_{\alpha \mu} D_{\beta} ; ~~~~~
[D_{\alpha}, D_{\beta}] =0
\eea
where
$ D_0 = \partial_0 \frac{sinh(ia\partial_0)}{ia\partial_0} + ia \partial^2_k \frac{e^{-ia\partial_0}}{2\varphi^2}$ and $ D_i = \partial_i \frac{e^{-ia\partial_0}}{\varphi} $.
Since non commutative field theories are expected to capture quantum gravity effects, it is naturally of interest and importance to study the construction and analysis of field theories, in particular gauge theories in non commutative space-time that have a correct commutative limit. It is also important that these deformed gauge theories allow one to calculate the deviations due to the non commutativity of space-time from their commutative counterparts in an easy and transparent way. Since the realisation approach is suited on the account of the above two requirements, we use it to construct $\kappa$-deformed gauge theory using generalisation of Feynman's approach.

\section{Wong's equation in $\kappa$-space-time}

The generators $I^a$, $a= 1, 2,.,n^2-1$ of $su(N)$ algebra satisfy the relation 
\be \label{isotopic spin}
[I^a,I^b] = i \hbar f^{abc} I^c.
\ee
For deriving the Wong's equation in $\kappa$-deformed space-time, we start from, 
\be\label{wongs1}
\frac{dI^a}{dt} = \frac{i}{\hbar} [\hat{H},I^a],
\ee 
where $\hat{H}$ is the Dirac Hamiltonian in $\kappa$-deformed space-time. The Dirac equation in $\kappa$-deformed space is\cite{sivakumar}
\be\label{Diraceq}
(i\gamma^{\mu} D_{\mu} - \frac{mc}{\hbar} )\psi =0,
\ee
where $\gamma^\mu$ are the Dirac matrices satisfying the relations $(\gamma^0)^2 = \mathbb{I},~~~\gamma^0 \gamma^i = \alpha^i ,~~[\gamma^\mu, \gamma^\nu]_+ = 2 \eta^{\mu \nu}$.

In our study, non commutative coordinates are expressed in terms of commutative variables \cite{SM} as,
\be \label{realisation}
\hat{x}_0 = x_0;~~~ \hat{x}_i = x_i(1-ia\partial_0)
\ee
where $x_0 = ct$, $x_i$ are the coordinates of the commutative space-time. Note that we have kept terms valid up to first order in $a$ only (see eq.(\ref{Rkappa})). There are different realisations possible for the $\hat{x}_\mu$ and these different choices correspond to different ordering \cite{SM}. All choices of realisations lead to non commutative corrections to observable quantities which are of the same order of magnitude\cite{AKP}. The components of the Dirac derivatives $D_\mu$ (valid up to first order in $a$ ) appearing in the above are \cite{SM,sivakumar}
\be \label{Derivatives}
D_{0} = \partial_{0} + \frac{ia}{2} \nabla^{2}; ~~~D_{i}= \partial_{i},
\ee
where $\nabla^2= \partial_i \partial_i$. Note the $D_\mu$ in eq.(\ref{Diraceq}) is not same as the gauge covariant derivative appearing in eq.(\ref{commutativeequation}). Note that $\partial_i$ and $\partial_0$ are the derivatives corresponding to the commutative coordinates. Here we have considered corrections valid only up to first order in the deformation parameter, $a$.
Substituting eq.(\ref{Derivatives}) in eq.(\ref{Diraceq}) gives
\be \label{Diraceq1}
\Big(i\gamma^{0}(\partial_{0} + \frac{ia}{2} \nabla^{2}) + i\gamma^{i}\partial_i-\frac{mc}{\hbar}\Big) \psi = 0.
\ee 
Using $p_{0} = i \hbar \partial_0$ and $p_i = i \hbar \partial_i$ and applying the minimal coupling prescription, i.e.,
\bea\label{minimal}
p_{0} \rightarrow p_0 - e A_0,~~
p_{i} \rightarrow p_{i}-e A_i,
\eea 
where $A_\mu$, $\mu = 0, 1, 2, 3$ are the components of the four-vector potential, and simplifying, we get the $\kappa$-deformed Dirac equation in the presence of an external non-abelian gauge field as,
\be \label{EQ1}
i \hbar \frac{\partial \psi}{\partial t} = \Big(-c \alpha^i (p_i -e A_i) + ec A_0 - \frac{ac}{2 \hbar} (p_i -  eA_i)^2 + mc^2 \beta \Big)  \psi,
\ee
where $\gamma_0$ is re-expressed as $\beta$.
From the above equation, we read off the $\kappa$-deformed Dirac Hamiltonian to be

\be\label{Hamiltonian}
\hat{H} = -c \alpha^i (p_i -e A_i) + ec A_0 - \frac{ac}{2 \hbar} (p_i -  eA_i)^2+ mc^2 \beta .
\ee 
Note that the $a$-dependent term in the above modified Hamiltonian is coming from the Dirac derivative.
We now use this modified Dirac Hamiltonian for finding Wong's equation in $\kappa$-deformed space-time.
Substituting Hamiltonian obtained in eq.(\ref{Hamiltonian}) in eq.(\ref{wongs1}) and using $A_\mu = A^a_\mu I^a$ gives 

\bea \nonumber
\frac{dI^a}{dt} &=& \frac{i}{\hbar} \bigg[ -c \alpha^i (p_i - e A^b_i I^b) + e c A^b_0 I^b -\frac{a c }{2 \hbar} (p_i - e A^b_i I^b)^2 + m c^2 \beta, ~I^a \bigg] \nonumber \\ 
&=& \frac{i}{\hbar} \bigg[ c e \alpha^i A^b_i [I^b, I^a] + e c A^b_0 [I^b, I^a] + \frac{a c e}{2 \hbar} \Big((p_i - e A_i) A^b_i [I^b, I^a] \nonumber\\&+& A^b_i [I^b, I^a] (p_i - e A_i) \Big) \bigg].
\eea

Using the ordering prescription where $x^\mu$  is always written to the left of $p^\mu$ for the third term in the above expression, we rewrite the above equation as
\bea \label{wong4} 
\frac{dI^a}{dt} &=& ec\alpha^if^{abc} A^b_i I^{c} + ec f^
{abc}A^b_0 I^c+ \frac{ace}{ \hbar} f^{abc} A^b_i I^c p_i - \frac{ace^2}{2 \hbar} f^{abc} A^b_i A^d_i[I^d I^c + I^c I^d].
\eea

Using Hamiltonian in eq.(\ref{Hamiltonian}) we now evaluate
\be \label{velocity}
c D_0 \hat{x}_j = \frac{i}{\hbar} [\hat{H}, \hat{x}_j],
\ee
Using the realisation for $D_0$ and $\hat{x}_j$ given in eq.(\ref{Derivatives}) and eq.(\ref{realisation}), respectively, we find 
\small{\be \label{lhs}
D_0 \hat{x}_j = (\partial_0 + \frac{ia}{2} \nabla^2) (x_j - \frac{a}{\hbar} x_j p_0),~~
=\partial_0 x_j (1- \frac{a p_0}{\hbar}). 
\ee}
Here we are considering corrections valid only up to first order in the deformation parameter $a$. The RHS of eq.(\ref{velocity}) is calculated using Hamiltonian in eq.(\ref{Hamiltonian}) as
\begin{eqnarray}
\frac{i}{\hbar} [\hat{H}, \hat{x}_j] = c\alpha_j - \frac{ac}{\hbar} (p_j - e A_j) - \frac{a}{\hbar} c \alpha_j p_0.
\end{eqnarray}
Equating these two equations, we get $c \alpha_i = c \partial_0 x_i + \frac{ac}{\hbar}(p_i -eA_i)$. Substituting this in eq.(\ref{wong4}), we find
\be \label{wongs}
\frac{dI^a}{d\tau} = e\dot{x}^{\mu} f^{abc} A^b_{\mu}I^c
- \frac{ace^2 }{2\hbar} f^{abc} A^{b}_i A^d_i (I^cI^d-I^dI^c),
\ee
where we have replaced $t$ with $\tau$ as we are in the relativistic case \cite{tanimura,MCL}.  
This is the $\kappa$-modified Wong's equation. Note that the non commutative terms in the modified Wong's equation are coming through the Dirac derivative (see eq.(\ref{Derivatives})) alone and not from the realisation in eq.(\ref{realisation}). Here we have also used the ordering prescription for $x$ and $p$.

It has been shown that the $\tau$-development is determined by the appropriate Hamiltonian \cite{fujii}, and this justifies the replacement of $t$ with $\tau$ that we have done. The role of $\tau$ has been further analysed in \cite{MCL}. Also note that $D_\mu$ introduced in eq.(\ref{generator}) are the generators of $\kappa$-deformed space-time translations. This justifies the use of $D_0$ on the LHS of eq.(\ref{velocity}), above.

\section{ Yang-Mills equations and Lagrangian in the $\kappa$-deformed space-time}
In $\kappa$-deformed space-time, the position coordinates obey the commutation relation given by eq.(\ref{kappa1}).
Differentiating eq.(\ref{kappa1}) with respect to the parameter $\tau$ one gets
\be \label{kappa2}
[\dot{\hat{x}}_{\mu}, \hat{x}_{\nu}] + [\hat{x}_{\mu},\dot{\hat{x}}_{\nu} ] = i (a_{\mu} \dot{\hat{x}}_{\nu} - a_{\nu} \dot{\hat{x}}_{\mu}).
\ee
Note that in the commutative limit, the RHS of the above equation identically vanishes, and the LHS is symmetric in $\mu$ and $\nu$. We introduce
\begin{equation}
[\hat{x}_{\mu},\dot{\hat{x}}_{\nu} ] =  -\frac{i \hbar}{m} \hat{\eta}_{\mu \nu}+i (a_{\mu} \dot{\hat{x}}_{\nu} - a_{\nu} \dot{\hat{x}}_{\mu}),
\end{equation}
which reduces to the correct commutative expression \cite{tanimura} in the limit $a \rightarrow 0$. Here $\hat{\eta}_{\mu \nu} = \eta_{\mu \nu} + a \delta \eta_{\mu \nu}$ is the $\kappa$-deformed metric  and it is known that the $\kappa$-deformed correction to the metric ($a \delta\eta_{\mu \nu}$) is not symmetric in $\mu$ and $\nu$ \cite{zuhair, HS}. Differentiating the above equation, we find
\begin{equation}
[\dot{\hat{x}}_{\mu}, \dot{\hat{x}}_{\nu}] = -[\hat{x}_{\mu}, \ddot{\hat{x}}_{\nu}]  -\frac{i\hbar}{m} \frac{d\hat{\eta}_{\mu \nu}}{d\tau} + i(a_{\mu} \ddot{\hat{x}}_{\nu} - a_{\nu} \ddot{\hat{x}}_{\mu}).
\end{equation}
Since LHS is anti-symmetric in $\mu$ and $\nu$ and RHS has symmetric and anti-symmetric terms in $\mu$ and $\nu$, we introduce two second rank tensors, $\hat{F}_{\mu \nu}$ and $\hat{S}_{\mu \nu}$ as
\begin{equation} \label{tensor}
m[\dot{\hat{x}}_{\mu}, \dot{\hat{x}}_{\nu}] = -\frac{i \hbar e}{m}\hat{F}_{\mu \nu} + \hat{S}_{\mu \nu}
\end{equation}
where $\hat{F}_{\mu \nu}$ is anti-symmetric in $\mu$ and $\nu$ and $\hat{S}_{\mu \nu} = S_{\mu \nu} + a\delta S_{\mu \nu}$ has one part which is symmetric and the other not symmetric in $\mu$ and $\nu$. Here note that  $\hat{S}_{\mu \nu} =-\frac{i\hbar}{m} \frac{d\hat{\eta}_{\mu \nu}}{d\tau} =   -\frac{ia\hbar}{m} \frac{d \delta \eta_{\mu \nu}}{d\tau} $ as $\eta_{\mu \nu}$ is a constant tensor.

Next consider the commutation relation
\be \label{D01}
[\hat{x}_0, I^a] = 0.
\ee
Differentiating eq.(\ref{D01}) with respect to $\tau$, we get
\be
[\dot{\hat{x}}_0, I^a] + [\hat{x}_0, \dot{I}^a] =0.
\ee
Substituting Wong's equation obtained in eq.(\ref{wongs}) for $\dot{I}^a$ in the second term and rearranging the above expression gives (where we have used $[\hat{x}_0, \dot{\hat{x}}_0] =-\frac{i\hbar}{m} \hat{\eta}_{00}$)
\be \label{commutation1}
[\dot{\hat{x}}_0, I^a] = \frac{i\hbar e}{m} f^{abc} A^b_0 I^c.
\ee
Here, as $[x_0, I^a]=0$, the $a$-dependent term in Wong's equation does not contribute to the above expression. Note that even if we would have introduced $\hat{A} = A+ a\delta A$ in the minimal coupling prescription (see eq.(\ref{minimal})), the corresponding $a$-dependent term of Wong's equation would not have changed eq.(\ref{commutation1}) (see discussion in the fourth paragraph of the conclusion). Next we find the commutation relation between $\dot{\hat{x}}_0$ and Lie algebra valued functions of $\kappa$-deformed coordinates. For this we start with an arbitrary function $\phi^a(\hat{x}_\mu)$ and consider its Taylor series expansion given by $\phi^a(\hat{x}_i) = \phi^a (x_i - \frac{a}{\hbar} x_i p_0) = \phi^a(x_i) - \frac{a}{\hbar} x_j \partial^j \phi^a p_0$, $\phi^a(\hat{x}_0) = \phi^a(x_0)$. Using this and the  realisation of non commutative coordinates in terms of commutative coordinates (see eq.(\ref{realisation}) and eq.(\ref{commutation1})), we get
\be \label{covaraint1}
[\dot{\hat{x}}_0, \phi^a(\hat{x}_{\mu})I^a] = \frac{i\hbar }{m} (\delta^{ac}\partial_0 - ef^{abc} A^{b}_0 ){\phi}^cI^a=\frac{i\hbar }{m} \tilde{D}^{ac}_0 \phi^{c} I^a
\ee
where $\tilde{D}^{ac}_0 \phi^c =( \delta^{ac}\partial_0 - ef^{abc} A^{b}_0)\phi^c $ is the gauge covariant derivative in $\kappa$-deformed space-time. Note that $\tilde{D}_0$ does not get any $a$ dependent correction up to first order in $a$.
Similarly, using
\be \label{commutation2}
[\dot{\hat{x}}_i, I^a] = \frac{i\hbar e}{m} f^{abc} A^b_i I^c - \frac{i a e}{m} f^{abc} A^b_i I^c p_0,
\ee
we find
\bea \label{covaraint2}
[\dot{\hat{x}}_i, \phi^a(\hat{x}_{\mu})I^a] &=&(1-\frac{a p_0} {\hbar})\frac{i\hbar }{m} \Big(\delta^{ac}\partial_i - ef^{abc} A^{b}_i \Big){\phi}^cI^a  =\frac{i\hbar }{m} \tilde{D}^{ac}_i \phi^{c} I^a
\eea
where $\tilde{D}_i^{ac} \phi^c  = (1-\frac{a p_0}{\hbar} ) (\delta^{ac}\partial_i - ef^{abc} A^{b}_i ){\phi}^c $ is the spatial part of the gauge covariant derivative in $\kappa$-deformed space-time. Here all the $a$-dependent terms are coming from the realisation of $\hat{x}_i$ (see eq.(\ref{realisation})). Thus, we obtain all the components of the $\kappa$-deformed gauge covariant derivative.
Using the Jacobi identity $[\dot{\hat{x}}_{\mu},[\dot{\hat{x}}_{\nu}, \dot{\hat{x}}_{\rho}]] + [\dot{\hat{x}}_{\nu},[\dot{\hat{x}}_{\rho}, \dot{\hat{x}}_{\mu}]] + [\dot{\hat{x}}_{\rho},[\dot{\hat{x}}_{\mu}, \dot{\hat{x}}_{\nu}]] =0$ and substituting eq.(\ref{tensor}) in the above, we find that the $\hat{F}_{\mu \nu}$ satisfies 
\be \label{YM1}
\epsilon^{0ijk} (\tilde{D}_i \hat{F}_{jk} + \tilde{D}_j \hat{F}_{ki} + \tilde{D}_k \hat{F}_{ij}) =0,
\ee 
\be \label{YM2}
\epsilon^{0ijk} (\tilde{D}_0 \hat{F}_{jk} + 2 \tilde{D}_j \hat{F}_{k0} ) =0,
\ee
where we used $[\dot{\hat{x}}^{\mu},\frac{d \delta \hat{\eta}^{\nu \rho}}{d\tau}] =0$ as the $\kappa$-deformed corrections to the metric is known to be independent of $\hat{x}^\mu$ and depend only on $p^\mu$ \cite{zuhair,HS}.
It remains to be shown that $\hat{F}_{ij}$ and $\hat{F}_{0i}$ are indeed $\kappa$-deformed Yang-Mills field strength. To show this, we start from the commutation relation between $\dot{\hat{x}}_{0}$ and $I^a$ given in eq.(\ref{commutation1}). Taking commutator of  eq.(\ref{commutation1}) with $\dot{\hat{x}}_{i}$ gives,
\be
0 =[\dot{\hat{x}}_i, [\dot{\hat{x}}_0, I^a]] - \frac{i\hbar e}{m} f^{abc} [\dot{\hat{x}}_i, A^b_0 I^c].
\ee
After re-express the first term as $[\dot{\hat{x}}_0,[\dot{\hat{x}}_i, I^a]] + [I^a,[\dot{\hat{x}}_{0},\dot{\hat{x}}_{i}]]$ using the Jacobi idenity and using eq.(\ref{realisation}), the above equation becomes
\bea 
0&=& [\dot{x}_0, [\dot{\hat{x}}_i, I^a]] -\frac{i \hbar e}{m^2} [I^a, \hat{F}^b_{0i} I^b] - \frac{i \hbar e}{m} f^{abc} \Big([\dot{x}_i - \frac{a}{\hbar} \dot{x}_i p_0,A^b_0 I^c]\Big).
\eea
In writing the second term in the above equation, we have used eq.(\ref{tensor}). Using eq.(\ref{commutation2}) and eq.(\ref{isotopic spin}), we get the expression for $\hat{F}^a_{0i}$ as
\begin{equation} \label{NCF0i}
\hat{F}^a_{0i}  = (\partial_0 A^a_i - \partial_i A^a_0 - e f^{abc} A^b_0 A^c_i) (1- \frac{a p_0}{\hbar}) = F^a_{0i} (1- \frac{a p_0}{\hbar}).
\end{equation}
Here $ F^a_{0i} = \partial_0 A^a_i - \partial_i A^a_0 - e f^{abc} A^b_0 A^c_i$, is the commutative ``Electric field''.

Next by taking commutator of eq.(\ref{commutation2}) with $\dot{\hat{x}}_j$, we get
\bea 
0 &=& [\dot{\hat{x}}_j, [\dot{\hat{x}}_i , I^a]] - \frac{i \hbar e}{m} f^{abc} ([\dot{\hat{x}}_j,A^b_i]I^c + A^b_i[\dot{\hat{x}}_j, I^c ]) + \frac{i a e}{m} f^{abc} ([\dot{\hat{x}}_j, A^b_i I^c p_0]).
\eea
As earlier, applying the Jacobi identity in the first term and using eq.(\ref{tensor}), we find 
\bea 
0&=& ~[\dot{\hat{x}}_i,[\dot{\hat{x}}_j, I^a] ] -\frac{i \hbar e}{m^2}  \hat{F}^b_{ij} [I^a,I^b] - \frac{i \hbar e}{m} f^{abc} ([\dot{\hat{x}}_j,A^b_i]I^c + A^b_i[\dot{\hat{x}}_j, I^c ]) +\frac{i a e}{m} f^{abc} ([\dot{\hat{x}}_j, A^b_i I^c p_0]).
\eea
Next, using eq.(\ref{commutation2}) and eq.(\ref{isotopic spin}) in the right-hand side of the above equation and simplifying, we get
\begin{equation}\label{NCFij}
\hat{F}_{ij}^a =  (\partial_i A^a_j - \partial_j A^a_i - e f^{abc} A^b_i A^c_j)(1-\frac{2a p_0}{\hbar})= F^a_{ij} (1-\frac{2ap_0}{\hbar}),
\end{equation}
where $ F^a_{ij} = \partial_i A^a_j - \partial_j A^a_i - e f^{abc} A^b_i A^c_j$ is the commutative ``magnetic'' field. Thus eq.(\ref{NCF0i}) and eq.(\ref{NCFij}) show that the eq.(\ref{YM1}) and eq.(\ref{YM2}) are indeed the $\kappa$-deformed Yang-Mills equations.

We observe that the second rank anti-symmetric tensor $\hat{F}_{\mu \nu}$ introduced in eq.(\ref{tensor}) is indeed the $\kappa$-deformed Yang-Mills field strength. Note that the $a$-dependent term of the Yang-Mills field strength is coming due to the non commutativity of the coordinates\footnote{From eq.(\ref{Rkappa}), it is tempting to assume that the non commutative correction to all orders would be an overall multiplication by $e^{-\frac{ap_0}{\hbar}}$ for $\hat{F}_{0i}$ and $e^{-\frac{2ap_0}{\hbar}}$ for $\hat{F}_{ij}$.}.
This form of $F_{\mu \nu}$ is compatible with
\be
([\tilde{D}_0, \tilde{D}_i]\phi)^a  = -e f^{abc} F^b_{0i} \phi^c(1-\frac{a p_0}{\hbar}) ,
\ee
valid up to first order in $a$. Similarly,
\be
([\tilde{D}_i, \tilde{D}_j]\phi)^a =-e f^{abc} F^b_{ij} \phi^c(1-\frac{2a p_0}{\hbar}).
\ee
We observe that the commutators of components of the $\kappa$-deformed gauge covariant derivatives exactly matches with the deformed Yang-Mills field strength we derived in $\kappa$-space-time (see eq.(\ref{NCF0i}) and eq.(\ref{NCFij})). We also note that the $a$-dependent corrections to $\hat{F}^a_{\mu \nu}$ are coming solely from the non commutativity of the coordinates, and not from the $a$-dependent corrections of Wong's equation.

In the non commutative space-time, we have found that $\hat{F}_{ij}^a = F^a_{ij} (1-\frac{2a p_0}{\hbar})$ and $\hat{F}^a_{0i} = F^a_{0i} (1-\frac{a p_0}{\hbar})$, where $F_{ij}$ and $F_{0i}$ are the commutative field strengths. Using these, we find the $SU(N)$ Yang-Mills Lagrangian as,
\begin{equation}\label{lag}
\hat{L} = -\frac{1}{2} (1- \frac{a p_0}{\hbar})^2 F^a_{0i} F^{a0i} - \frac{1}{4} (1- \frac{2a p_0}{\hbar})^2 F^a_{ij} F^{aij}.
\end{equation}
Here $p^0$ is the energy scale of the non commutative space-time, that is, the energy of the probe that sees the non commutativity (and its effects).

This Lagrangian is invariant under the usual $SU(N)$ gauge transformations. We have already derived two of the Yang-Mills equations (see eq.(\ref{YM1}) and eq.(\ref{YM2})). The remaining two equations follow as the Euler-Lagrange equation from the above Lagrangian, and they are
\be
(1-\frac{2a p_0}{\hbar})( D_i F^{0i})^a = 0,~~(D_0 F^{i0})^a +(1-\frac{2a p_0}{\hbar})( D_j F^{ij})^a = 0.
\ee
In the limit $a\rightarrow 0$, above equations reduce to the commutative result.
It is clear that under the gauge transformation of the vector field, $A^a_{\mu} \rightarrow A^a_{\mu}+ (\delta^{ac}\partial_\mu -ef^{abc} A^b_\mu)\alpha^c(x)$, the Lagrangian given in eq.(\ref{lag}) is invariant, showing the invariance of $\kappa$-deformed Yang-Mills field theory under usual $SU(N)$ transformation.

\section{Lorentz force equation in presence of $SU(N)$ gauge fields}

To derive the force experienced by an isospin carrying particle in the presence of Yang-Mills field in $\kappa$-space-time, we start with $[\hat{x}_i, \hat{F}_0]$ where $\hat{F}_0$ is the zeroth-component of the 4-vector force. We first do a Taylor series expansion of $\hat{F}_0$, considering it to be a function of $\hat{x}_\mu$ and $\dot{\hat{x}}_\mu$ given in eq.(\ref{realisation}). We then use this Taylor series expansion of $\hat{F}_0 $ keeping terms to the first order in $a$, i.e., $\hat{F}_0 = F_0 - \frac{a}{\hbar} (x_j \partial^j F_0 p_0 + \dot{x}_j \frac{\partial F_0}{\partial \dot{x}_j} p_0)$ in the above equation. For $\hat{x}_i$, we use the realisation given in eq.(\ref{realisation}). Thus we find
\bea 
[\hat{x}_i, \hat{F}_0] &=& [x_i -\frac{a}{\hbar} x_i p_0,~F_0 - \frac{a}{\hbar} (x_j \partial^j F_0 p_0 + \dot{x}_j \frac{\partial F_0}{\partial \dot{x}_j} p_0)].
\eea
Keeping corrections valid only up to first order in the deformation parameter $a$, we find
\bea \label{F01} 
[\hat{x}_i, \hat{F}_0] &=& \frac{i \hbar e}{m} F_{i0} - \frac{a}{\hbar} \Big( {\frac{i \hbar e}{m}} F_{i0} p_0 - \frac{i \hbar}{m} x_j \frac{\partial (\partial^j F_0)}{\partial \dot{x}^i} p_0 -\frac{i\hbar}{m}\frac{\partial F_0}{\partial\dot{x}^i} p_0 -\frac{i \hbar}{m} \dot{x}_j \frac{\partial (\frac{\partial F_0}{\partial \dot{x}_j})}{\partial \dot{x}^i} p_0\Big).
\eea
Considering $\hat{F}_0 = F_0 + a \delta F_0$, where $\delta F_0$ is the modified part of force in the $\kappa$- deformed space-time, we find the commutator,
\bea \label{F02}
[\hat{x}_i, \hat{F}_0] = [x_i - \frac{a}{\hbar} x_i p_0,~F_0 + a \delta F_0].
\eea
From eq.(\ref{F01}) and eq.(\ref{F02}), we equate the commutative and non commutative parts separately.
Equating commutative part gives, $\frac{i \hbar e}{m} F_{i0} = -\frac{i \hbar}{m} \frac{\partial F_0}{\partial \dot{x}^i}$, which on integration gives,
\be \label{F0commutative}
F_0 = e F_{0i} \dot{x}^i + G_0(x),
\ee
where $G_0(x)$, is a function of $x$ alone.
Now equating the non commutative part of the equations eq.(\ref{F01}) and eq.(\ref{F02}) gives,
\bea \label{Fc} 
-\frac{i \hbar}{m} \frac{\partial \delta F_0}{\partial \dot{x}^i} &=& - \frac{1}{\hbar} \bigg[{\frac{i \hbar e}{m}} F_{i0} p_0 - \frac{i \hbar}{m} x_j \frac{\partial (\partial^j F_0)}{\partial \dot{x}^i} p_0 - \frac{i \hbar}{m} \dot{x}_j \frac{\partial (\frac{\partial F_0}{\partial \dot{x}_j})}{\partial \dot{x}^i} p_0\bigg].
\eea
By substituting $F_0$ from the commutative result (eq.(\ref{F0commutative})) and integrating, we get
\be \label{Fnc}
\delta F_0 = \frac{1}{\hbar} [eF_{i0}p_0 - ex_j \partial^j F_{0i} p_0 ]\dot{x}^i + \delta {G}_0(x).
\ee
Using eq.(\ref{F0commutative}) and  eq.(\ref{Fnc}), we find
\be \label{force0 final}
\hat{F}_0 = e F_{0i} \dot{x}^i  (1-\frac{a}{\hbar}p_0)- \frac{a e}{\hbar} x_j \partial^j F_{0i} \dot{x}^i p_0 + \hat{G}_0(x),
\ee
where $\hat{G}_0 = G_0 + a\delta {G}_0$.
Similarly, for deriving the expression for $\hat{F}_i$, consider
\be \label{Fi}
[\hat{x}_i, \hat{F}_j] = \bigg[x_i - \frac{a}{\hbar} x_i p_0, ~ F_j -\frac{a}{\hbar} (x_k \partial^k F_j p_0 + \dot{x}_k \frac{\partial F_j }{\partial \dot{x}_k} p_0)\bigg].
\ee
Here we have used the realisation of non commutative coordinates in terms of commutative coordinates, eq.(\ref{realisation}) and used Taylor series expansion of $\hat{F}_i$, valid up to first order in $a$.
Using $[x_i,F_j] =\frac{i\hbar e}{m} F_{ij}$ we get,
\bea \label{Fi1} 
[\hat{x}_i, \hat{F}_j] &=& \frac{i \hbar e}{m} F_{ij}-\frac{a}{\hbar}\Big( {\frac{i \hbar e}{m}} F_{ij} p_0 - \frac{i \hbar}{m} x_k \frac{\partial (\partial^k F_j)}{\partial \dot{x}^i} p_0 - \frac{i \hbar }{m} \frac{\partial F_j}{\partial \dot{x}^i} p_0- \frac{i \hbar}{m} \dot{x}_k \frac{\partial (\frac{\partial F_j}{\partial \dot{x}_k})}{\partial \dot{x}^i} p_0\Big).
\eea
Taking $ \hat{F}_j = {F}_j + a \delta F_j$, we find
\be \label{Fi2}
[\hat{x}_i, \hat{F}_j] = [x_i - \frac{a}{\hbar} x_i p_0,~ F_j + a \delta {F}_j],
\ee
and equating the commutative part of eq.(\ref{Fi1}) and eq.(\ref{Fi2}) gives, $\frac{i \hbar e}{m} F_{ij} = -\frac{i \hbar}{m} \frac{\partial F_j}{\partial \dot{x}^i}$,
which on integration gives
\be
F_j = e F_{ji} \dot{x}^i + G_j(x).
\ee
From the non commutative part of eq.(\ref{Fi1}) and eq.(\ref{Fi2}), we get
\be
-\frac{i \hbar}{m} \frac{\partial \delta {F}_j}{\partial \dot{x}^i} =  \frac{i}{m} \bigg[ x_k \frac{\partial (\partial^k F_j)}{\partial \dot{x}^i} p_0  + \frac{\partial F_j}{\partial \dot{x}^i} p_0+\dot{x}_k \frac{\partial (\frac{\partial F_j}{\partial \dot{x}_k})}{\partial \dot{x}^i} p_0 \bigg].
\ee
Using the commutative expression for $F_j = e F_{ji}\dot{x}^i + e F_{j0} \dot{x}^0 + G_j$ \cite{tanimura}, we get, $\frac{\partial \delta {F}_j}{\partial \dot{x}^i}= -\frac{1}{\hbar} (ex_k \partial^k F_{ji} p_0 + e F_{ji}p_0 )$.
Integrating the above equation gives,
\be
\delta {F}_j = -\frac{1}{\hbar} (ex_k \partial^k F_{ji} p_0 \dot{x}^i + e F_{ji}p_0 \dot{x}^i) + \delta {G}_j(x).
\ee
Thus, we get
\be \label{forcei1}
\hat{F}_i = e F_{ij} \dot{x}^j (1-\frac{a p_0}{\hbar}) -\frac{a}{\hbar} e x_k \partial^k F_{ij}  \dot{x}^j p_0 + \hat{G_i}(x). 
\ee
Here we are considering terms valid only up to first order in $a$ and thus $\hat{G}_i = {G}_i + a \delta {G}_i$. To find the complete expression of $\hat{F}_i$, we consider $[\hat{x}_0, \hat{F}_i]$ also, i.e.,
\bea 
[\hat{x}_0, \hat{F}_i] &=& [x_0, F_i -\frac{a}{\hbar} (x_k \partial^k F_i p_0 + \dot{x}_k \frac{\partial F_i }{\partial \dot{x}_k} p_0)] \nonumber \\
&=& \frac{i \hbar e}{m} F_{0i}+\frac{a}{\hbar}\Big( \frac{i \hbar}{m} x_k \frac{\partial (\partial^k F_i)}{\partial \dot{x}_0} p_0 + i \hbar x_k \partial^k F_i + \frac{i \hbar}{m} \dot{x}_k \frac{\partial (\frac{\partial F_i}{\partial \dot{x}_k})}{\partial \dot{x}_0} p_0 + i \hbar \dot{x}_k \frac{\partial F_i}{\partial \dot{x}_k}\Big).
\eea
As earlier, we consider $\hat{F}_i = F_i + a \delta F_i$, and find
\be
[\hat{x}_0, \hat{F}_i] = [x_0, F_i + a \delta F_i] = -\frac{i \hbar}{m} \frac{\partial F_i}{\partial \dot{x}_0} - \frac{i \hbar a}{m} \frac{\partial \delta F_i}{\partial \dot{x}_0}.
\ee
On equating the commutative and non commutative part, we get
\be \label{Ficommutative}
F_i = e F_{i0} \dot{x}_0 + G_i(x),
\ee
\bea \label{forcei2} 
\delta F_i &=& -\frac{1}{\hbar} \Big(e x_k \partial^k F_{i0} (p_0 \dot{x}_0 +m\frac{\dot{x}_0^2}{2} )  + me x_k \partial^k F_{ij} \dot{x}^j\dot{x}_0 \nonumber + m x_k \partial^k G_i \dot{x}_0 + me \dot{x}_k F^k_i \dot{x}_0\Big) + \delta {G}_i(x).
\eea
Combining eq.(\ref{forcei1}), eq.(\ref{Ficommutative}) and eq.(\ref{forcei2}) we get the expression for force in $\kappa$-deformed space-time as,

\bea \label{forcei final} \nonumber
 \hat{F}_i &=& e F_{i\mu} \dot{x}^\mu + \hat{G_i}(x) -\frac{a }{\hbar} \Big( e F_{ij} \dot{x}^j p_0 + e x_k \partial^k F_{i\mu} \dot{x}^\mu p_0 
\nonumber \\&+& m e x_k \partial^k F_{ij} \dot{x}^j \dot{x}_0 + m e x_k \partial^k F_{i0} \frac{\dot{x}_0^2}{2} + m x_k \partial^k G_i \dot{x}_0 + m e \dot{x}_k F^k_i \dot{x}_0 \Big).
\eea
Eq.(\ref{force0 final}) and eq.(\ref{forcei final}) give the components of force experienced by an isospin carrying particle in the presence of Yang-Mills field in $\kappa$-deformed space-time. Both these equations reduce to well-known results in the commutative limit (i.e., $ \lim {a \to 0}$).

\section{Conclusion}

We have constructed $SU(N)$ Yang-Mills theory in $\kappa$-deformed space-time, by generalising Feynman's approach. This should be contrasted with earlier studies where non-commutative gauge theories constructed all were invariant under $U(N)$ gauge group and not under $SU(N)$ gauge group. $SU(N)$ Yang-Mills theory developed here makes generalisation of standard model to $\kappa$-space-time amenable.

For deriving the $\kappa$-deformed $SU(N)$ Yang-Mills theory, we have first constructed Wong's equation in $\kappa$-deformed space-time. We have also obtained the force equation for a particle having isospin in the presence of a non-abelian gauge field. We have then constructed the Lagrangian for $\kappa$-deformed Yang-Mills theory and showed its invariance under $SU(N)$ gauge transformations. Thus, we see that the $\kappa$-deformed Yang-Mills theory obtained has the same gauge symmetry as its commutative counterpart. In our study, we have considered corrections valid only up to first order in the deformation parameter $a$.

In our study, by considering the symmetry properties of commutators between velocities, we have introduced a second-rank, anti-symmetric tensor, as well as another second-rank tensor; first reduces 
to the $SU(N)$ field strength \cite{tanimura} while the second vanishes in the commutative limit. We have then shown that the anti-symmetric second rank tensor is the Yang-Mills field strength tensor in $\kappa$-deformed space-time. Using the Jacobi identity repeatedly, we have showed that this second-rank, anti-symmetric tensor obeys homogeneous Yang-Mills field equations. In evaluating the Jacobi identity, we need commutation relation between Lie algebra-valued functions and $\kappa$-deformed velocities, which in turn leads to the commutation relation between generators of the Lie algebra and the velocities. This is obtained by using Wong's equation in $\kappa$-space-time, which we constructed using the $\kappa$-deformed Dirac Hamiltonian. Then by repeated use of Jacobi identity, we have derived two of the homogeneous Yang-Mills field equations in $\kappa$-deformed space-time.

Note that the $\kappa$-deformed Dirac equation is constructed \cite{sivakumar} using the Dirac derivatives \cite{SM} which are written in terms of derivatives with respect to the commutative space-time coordinates, their derivatives, and the deformation parameter. Thus, the $\kappa$-Dirac equation in momentum space will depend only on the commutative momentum $p_0$, $p_i$. Therefore, the minimal coupling method used to bring in the gauge interaction allows us to introduce only gauge field living in the commutative space-time. Here, if by brute force we use $\hat{A}^\mu (\hat{x})$, eq.(\ref{commutation1}) will have $a$-dependent term in the RHS while LHS is independent of any, leading to inconsistency. This strengthens our use of $A(x)$ in implementing minimal coupling in eq.(\ref{EQ1}). By minimal prescription, from eq.(\ref{Diraceq1}), we get eq.(\ref{EQ1}). In deriving this, we have replaced $p_0$ and $p_i$ with $p_0-eA_0$ and $p_{i}-eA_i$, where $A^a_0 I^a$ and $A^a_i I^a$ are the components of the non-abelian gauge field defined on the commutative space-time. This Dirac equation reduces to the correct commutative limit. It may be possible that one can add more $a$-dependent terms involving the gauge field such that the equation reduces to the correct commutative limit. But in the absence of any constructive procedure to introduce such terms, we do not venture into this. If we would have introduced gauge field which is a function of non commutative coordinates in implementing minimal coupling, we would have had more $a$-dependent terms with gauge field (actually its derivatives) in the deformed Dirac equation coming from the Taylor series expansion of $\hat{A}_\mu(\hat{x})$ using eq.(\ref{realisation}) for $\hat{x}_\mu$. But this is not the correct procedure as we have used Dirac derivatives (eq.(\ref{Derivatives})) which depend on $a$ and thus all contributions due to non commutativity have been included (particularly see the third term on RHS of eq.(\ref{EQ1})). Further, as stated, after eq.(\ref{commutation1}), $a$-dependent terms of the Wong's equation do not contribute to the RHS of eq.(\ref{commutation1}). Only if the $\kappa$-Dirac Hamiltonian has terms depending on $(ax^0) F(A)$, where $F(A)$ is a function of $A$, the RHS of eq.(\ref{commutation1}) would be modified. As $\hat{x}_0 =x_0$, such a term will not come by Taylor expansion of $\hat{A}(\hat{x})$ (see discussions after eq.(\ref{commutation1})). This clearly shows that the minimal coupling procedure used in eq.(\ref{minimal}) is on strong grounds.

Note that in the $\kappa$-space-time, it is the time coordinate that does not commute with space coordinates, while space coordinates commute among themselves. Also, here the commutation relation depends on the space coordinates (see eq.(\ref{kappa1})). This should be contrasted with the Moyal space-time where the commutators between coordinates are proportional to a constant (antisymmetric) tensor and thus the no-go theorem obtained in Moyal space-time \cite{chaichian} is not directly applicable to our study. For a non commutative space-time whose coordinates obey $[\hat{x}^{\mu}, \hat{x}^{\nu}] = i \theta^{\mu \nu}(\hat{x})$, gauge theory was constructed in \cite{calmet} and it was argued that the closure of the gauge transformation to all order in $\theta(\hat{x})$ will necessitates the fields to be valued in the enveloping algebra rather than Lie algebra. It is not clear whether neglecting second and higher-order terms in the deformation parameter $a$ is the reason for our gauge field to be Lie algebra valued, and further study is needed to clarify this. But the Feynman approach and its covariant generalisation do not, a priori, demand the existence of a gauge field, and hence it is not clear how including higher order corrections will hinder this construction of the $SU(N)$ Yang-Mills equation obtained here. We plan to study this issue in detail and will be reported in future. 

We have obtained the explicit expression for $\hat{F}_{0i}$ and $\hat{F}_{ij}$ and found their $a$-dependent corrections. We find the correction to $\hat{F}_{0i}$, $\delta \hat{F}_{0i} = -\frac{ap_0}{\hbar} F_{0i}$ whereas that of $\hat{F}_{ij}$, $\delta \hat{F}_{ij} = -\frac{2ap_0}{\hbar} F_{ij}$. We have also obtained the explicit expression valid up to first order in $a$ for the gauge covariant derivative $\tilde{D}_0$ and $\tilde{D}_i$ and observed that $\tilde{D}_0$ does not get any $a$-dependent correction while $\tilde{D}_i$ does have $a$-dependent correction as $-\frac{a p_0}{\hbar}  (\delta^{ac}\partial_i - ef^{abc} A^{b}_i ){\phi}^c $.

We have also showed that, as in the commutative case, the commutators of the components of the $\kappa$-deformed gauge covariant derivative, $([\tilde{D}_0, \tilde{D}_i]\phi)^a $ and $([\tilde{D}_i, \tilde{D}_j]\phi)^a$  match with the Yang-Mills field strength tensors $\hat{F}_{0i}$ and $\hat{F}_{ij}$, respectively, derived by explicit calculations. We have constructed the Lagrangian describing Yang-Mills theory in $\kappa$-space-time using the field strength tensor, and two of the Yang-Mills field equations are derived by finding the corresponding Euler-Lagrange equations.

Note that the $\kappa$-deformed Yang-Mills Lagrangian is written in terms of $\hat{F}_{ij}^a$ and $\hat{F}^a_{0i}$. Since  $\hat{F}_{ij}^a = F^a_{ij} (1-\frac{2a p_0}{\hbar})$ and $\hat{F}^a_{0i} = F^a_{0i} (1-\frac{a p_0}{\hbar})$ where $F^a_{ij}$ and $F^a_{0i}$ are the components of the field strength in the commutative space-time, we see that they are covariant under the usual $SU(N)$ gauge transformation i.e., under $A^a_{\mu} \to A^a_{\mu} + (\delta^{ac}\partial_\mu - ef^{abc} A^{b}_\mu ){\lambda}^c $. Note that the gauge transformation is given in terms of the usual, commutative space-time covariant derivative. Thus, our $\kappa$-deformed Yang-Mills Lagrangian is invariant under the usual $SU(N)$ transformation.  We have also verified that in the commutative limit, i.e., $a \rightarrow 0$, we get the commutative results, as expected.
 
In \cite{kurkov}, the U(1) gauge covariant field strength ($F_{\mu \nu}$) constructed in $\kappa$-space-time has quadratic and cubic terms in the gauge field even in first order in the deformation parameter. Though in this work only first-order terms in the non commutative parameter in the commutators between the space-time coordinates are included, the gauge transformations and field strength have all order terms in the non commutative parameter. Thus, the U(1) action constructed has terms higher than quadratic in fields. All higher-order terms in fields are $a$ dependent and vanish when $a \rightarrow 0$. This should be contrasted with the $SU(N)$ gauge field strength we obtained in eq.(\ref{NCF0i}) and eq.(\ref{NCFij}) in $\kappa$-deformed space-time. Here the non commutative correction appears through an overall multiplicative factor, i.e., $\hat{F}_{ij}^a = F^a_{ij} (1-\frac{2a p_0}{\hbar})$ and $\hat{F}^a_{0i} = F^a_{0i} (1-\frac{a p_0}{\hbar})$, where $F^{a}_{0i}$ and $F^{a}_{ij}$ are components of field strength as in the commutative space-time. It is the fact that the $\kappa$-deformation modifying the components of the field strength $\hat{F}^{a}_{0i}$ and $\hat{F}^{a}_{ij}$ only by overall multiplicative factor is responsible for keeping the $SU(N)$ invariance intact in the Lagrangian obtained in eq.(\ref{lag}).
Since the effect of the non commutativity of space-time on the
'electric' and 'magnetic' sectors of the deformed gauge theory constructed here is different, it is of interest to study its implications, in particular when coupled to matter fields. Work along this direction is in progress and will be reported separately.

\section{Acknowledgement}
BR thanks DST-INSPIRE for support through the INSPIRE fellowship (IF220179).

\end{document}